\documentclass[a4paper,11pt]{article}
\usepackage{pos}
\usepackage{braket}
\usepackage{caption}
\usepackage{amsmath}
\usepackage{slashed}
\allowdisplaybreaks

\title{{\normalfont\footnotesize{MSUHEP-22-024, ZU-TH 32/22}}\\[1ex]
Renormalization of twist-two operators in QCD and its application to singlet splitting functions}
 \ShortTitle{Renormalization of twist-two operators in QCD}

\author[a]{Thomas Gehrmann}
\author[b]{Andreas von Manteuffel}
\author*[a,b]{Tong-Zhi Yang}

\affiliation[a]{Department of Physics, University of Z\"urich, CH-8057 Z\"urich, Switzerland}

\affiliation[b]{Department of Physics and Astronomy, Michigan State University, East Lansing, Michigan 48824, USA}

\emailAdd{thomas.gehrmann@uzh.ch}
\emailAdd{vmante@msu.edu}
\emailAdd{toyang@physik.uzh.ch}

\abstract{Splitting functions govern the scale evolution of parton distribution functions. Through a Mellin transformation, they are related to anomalous dimensions of twist-two operators in the operator product expansion. We study off-shell operator matrix element, where the physical operators mix under renormalization with other gauge-variant operators of the same quantum numbers. We devise a new method to systematically extract the Feynman rules resulting from those operators without knowing the operators themselves. As a first application of the new approach, we independently reproduce the well-known three-loop singlet splitting functions obtained from computations of on-shell quantities.}

\FullConference{%
  Loops and Legs in Quantum Field Theory - LL2022,\\
  25-30 April, 2022\\
  Ettal, Germany
}

\begin{document}
\maketitle

\section{Introduction}
\label{sec:introduction}
With the accumulated collision data of Run 2 at the Large Hadron Collider (LHC), and especially the data that will be collected in the High-Luminosity phase of LHC (HL-LHC), the experimental uncertainty keeps decreasing for many benchmark processes, motivating theory predictions with correspondingly high precision. Currently, several benchmark computations in  Quantum Chromodynamics (QCD) are available at next-to-next-to-next-to-leading order (N$^3$LO) for inclusive cross sections~\cite{Anastasiou:2015vya,Anastasiou:2016cez,Mistlberger:2018etf,Duhr:2020seh,Duhr:2020sdp}, and for more differential distributions~\cite{Cieri:2018oms,Chen:2021isd,Chen:2021vtu,Chen:2022cgv,Chen:2022lwc}. The resulting predictions are based on NNLO parton distribution functions (PDFs) and therefore carry a residual uncertainty due to the missing N$^3$LO PDFs. The evolution of N$^3$LO PDFs requires the N$^3$LO (four-loop) splitting functions, which are currently still not completely known. Furthermore, the four-loop splitting functions are also important ingredients to complete N$^4$LL resummations for Drell-Yan and single Higgs production. 

Given the importance of splitting functions, they are widely studied in the literature. Many fruitful results are extracted from deep-inelastic scattering (DIS) or DIS-like processes with one-loop~\cite{Altarelli:1977zs}, two-loop~\cite{Curci:1980uw,Furmanski:1980cm}, three-loop~\cite{Moch:2004pa,Vogt:2004mw}, and low moments four-loop results~\cite{Moch:2021qrk} available. The splitting function results to three-loop have been verified multiple times, from hadronic cross sections~\cite{Mistlberger:2018etf,Duhr:2020seh} or from operator matrix elements (OMEs) which we will elaborate upon in more detail below.

As one of the most efficient methods for the determination of splitting functions~\cite{Gross:1974cs}, the OME method has been studied for a long time. The OMEs are defined as matrix elements of local operators,
\begin{align}
A^{}_{ij} = \braket{j(p)|O_i|j(p)}^{},\qquad  i, j = q \text{ or } g \,.
\label{eq:ome}
\end{align}
The operators we consider in the following are the so-called leading twist (twist-two) contributions in the operator product expansion that encode the collinear physics at leading power. According to the flavour group, the twist-two operators are divided into non-singlet and singlet parts. The non-singlet operators of spin-$n$ are given by
\begin{equation}
\label{eq:nonsingletOP}
O^{}_{q,k} = \frac{1}{2}  \big[ \bar{\psi}_{i} \slashed{\Delta} (\Delta \cdot D)_{ij}^{n-1}\frac{\lambda_k}{2} \psi_{j}  \big]\,,  
\end{equation}  
where $\lambda_k$ is a diagonal generator of the flavour group. The singlet quark and gluon operators are obtained as
\begin{align}
\label{eq:singletOP}
O^{}_{q} &= \frac{1}{2}  \big[ \bar{\psi}_{i} \slashed{\Delta} (\Delta \cdot D)_{ij}^{n-1} \psi_{j}  \big] \,, & O^{}_g &= \frac{1}{2} \big[ \Delta_{\mu_1} G^{\mu_1}_{a, \mu}  (\Delta \cdot D)_{ab}^{n-2}  \Delta_{\mu_n} G^{\mu_n \mu}_{b}   \big] \,,
\end{align}
where the covariant derivative in quark and gluon operators are defined by 
\begin{align}
\label{eq:derivFund}
(D_\mu)_{ij} &= \partial_\mu \delta_{ij} - i g_s (T^a)_{ij} A^a_\mu \,, &
D^{ab}_\mu &= \partial_\mu \delta^{ab} - g_s f^{abc} A^c_\mu \,,
\end{align}
respectively.
The non-singlet operator $O_{q,k}$ is distinguished from singlet operators by the quark flavor and is multiplicatively renormalized, 
\begin{equation}
\label{eq:NonsingletRe}
O^{\text{R}}_{q,k}  = Z^{\text{ns}} O^{\text{B}}_{q,k} \,.
\end{equation} 
The two singlet operators have the same quantum number and therefore mix under renormalization,
\begin{align}
\label{eq:mixingOqOg}
\left( \begin{array}{c} 
O_q \\
O_g
\end{array} \right)^{\text{R}} =  \left( \begin{array}{cc} 
Z_{qq} & Z_{qg} \\
Z_{gq} & Z_{gg} 
\end{array} \right)   \left( \begin{array}{c} 
O_q \\
O_g 
\end{array} \right)^{\text{B}}\,. 
\end{align}
The anomalous dimensions $\gamma_{ij}$ of the twist-two operators are related to the splitting functions through a Mellin transformation 
\begin{align}
 \gamma_{ij}(n)=-\int_{0}^{1}dz \; z^{n-1}P_{i j}(z) \,,
\end{align}
and the coefficients of the perturbative expansion $\gamma_{ij} = \sum_{l=0}^\infty \gamma_{ij}^{(l)} (\alpha_s/(4\pi))^{l+1}$ can be extracted from the single pole of the corresponding renormalization factors,
\begin{align}
\label{eq:ZfactorIntermsofGamma}
Z_{ij} &= \delta_{ij} + \left(\frac{\alpha_s}{4 \pi}\right) \frac{\gamma^{(0)}_{ij}}{\epsilon}+\left(\frac{\alpha_s}{4 \pi}\right)^2 \bigg( \frac{\gamma_{ij}^{(1)}}{2 \epsilon} + \frac{1}{2 \epsilon^2} \big[ -\beta_0 \gamma_{ij}^{(0)} + \sum_{k=q,\,g} \gamma^{(0)}_{ik} \gamma^{(0)}_{kj}  \big] \bigg) \nonumber 
\\
& + \left(\frac{\alpha_s}{4 \pi}\right)^3 \bigg( \frac{1}{3 \epsilon} \gamma_{ij}^{(2)} + \frac{1}{6 \epsilon^2} \big[ -2 \beta_1 \gamma_{ij}^{(0)} - 2 \beta_0 \gamma_{ij}^{(1)} +2 \sum_{k=q,\,g} \gamma^{(1)}_{ik} \gamma^{(0)}_{kj}  +\sum_{k=q,\,g} \gamma^{(0)}_{ik} \gamma^{(1)}_{kj}    \big] \nonumber 
\\
& \quad + \frac{1}{6\epsilon^3} \big[ 2 \beta_0^2 \gamma_{ij}^{(0)}  - 3 \beta_0 \sum_k \gamma_{ik}^{(0)} \gamma_{kj}^{(0)} + \sum_{k=q,\,g} \sum_{l=q,\,g} \gamma^{(0)}_{ik} \gamma^{(0)}_{kl} \gamma^{(0)}_{lj}   \big] \bigg)+ \mathcal{O}(\alpha_s^4)\,,
\end{align}
where $\alpha_s = g_s^2/(4 \pi)$, $g_s$ is the strong coupling constant, and $\epsilon = (4-d)/2$ is the dimensional regulator.

Depending on the nature of the external parton states in Eq.~(\ref{eq:ome}), one 
distinguishes on-shell and off-shell OMEs. The renormalization procedure in Eq.~\eqref{eq:NonsingletRe} and Eq.~\eqref{eq:mixingOqOg} remains valid only 
for on-shell OMEs. However, at least one-mass scale is needed to keep the on-shell OMEs non-vanishing in dimensional regularization. In the literature, three approaches were proposed to generate the mass scale. The first one is to consider an operator insertion with non-zero momentum transfer and the one-loop case was studied in Ref.~\cite{Harris:1994tp}. The method introduces not only one mass scale but also an external momentum resulting in much more scalar products, such that the calculation becomes very involved at higher orders. The second method introduces an internal mass scale and was widely used to study the 3-loop heavy flavour quark contributions in DIS~\cite{Ablinger:2014nga,Ablinger:2014vwa,Ablinger:2017tan,Behring:2019tus}. The third method introduces the mass scale by imposing phase-space constraints and was used to compute, for example, three-loop beam functions~\cite{Luo:2019szz,Ebert:2020yqt,Ebert:2020unb,Luo:2020epw}. As a by-product, these calculations based on the second and the third method also confirm the splitting functions to three loops independently.

Regarding the off-shell OMEs method, there is no need to introduce other scales since the off-shellness of the external particle sets the mass scale. From the point of view of computational efficiency, the computation of purely massless off-shell OMEs is preferable over on-shell OMEs with phase-space constraints or with an internal mass. The massless off-shell OMEs are however non-physical objects since the virtuality of external quark or gluon states should be zero in massless QCD. For the non-singlet case, the renormalization in Eq.~\eqref{eq:NonsingletRe} is still correct, since only a single operator exists for a specified quark flavour and it can not mix with other operators. As a result, computations of non-singlet splitting functions based on the off-shell OMEs method are very advanced. Historically, both the one-loop~\cite{Gross:1973ju} and two-loop~\cite{Floratos:1977au} non-singlet splitting functions were first calculated using this method.
The three-loop non-singlet splitting functions were rederived with this method in Ref.~\cite{Blumlein:2021enk,Blumlein:2022kqz}.
At four loops, partial results for the non-singlet splitting functions have been obtained in Ref.~\cite{Moch:2017uml}.
In particular, the authors reconstructed the full-$n$ leading-color results from fixed Mellin moments up to $n=20$.
Together with Mellin moments up to $n=16$ for the full-color contribution, these results constitute the state-of-the-art precision for non-singlet splitting functions.

For the singlet case, three-loop calculations based on the massless off-shell OME method are available only for the polarized splitting functions~\cite{Moch:2014sna,Blumlein:2021ryt}.
Calculations for the unpolarized singlet splitting functions, however, are much less advanced with this method.
This is mainly because the renormalization procedure in Eq.~\eqref{eq:mixingOqOg} is no longer valid in the sense that the operators $O_q$ and $O_g$ will also mix with some unknown a priori gauge-variant (alien) operators. The new mixing was first pointed out by Gross and Wilczek in 1974~\cite{Gross:1974cs} when the first one-loop singlet splitting function was calculated. In the seminal work of Dixon and Taylor~\cite{Dixon:1974ss} the set of gauge-variant operators relevant at order $g_s$ was constructed explicitly. These results enabled subsequently the determination of the correct singlet renormalization at two loops~\cite{Hamberg:1991qt}, thereby resolving earlier inconsistencies~\cite{Floratos:1978ny,Gonzalez-Arroyo:1979qht}. All-order renormalization conditions for the gauge-invariant operators $O_q$ and $O_g$ were derived by Joglekar and Lee~\cite{Joglekar:1975nu}, confirming the results of~\cite{Dixon:1974ss} but not providing a procedure for the construction of gauge-variant operators at higher orders in $g_s$.

Most recently, Falcioni and Herzog~\cite{Falcioni:2022fdm,Falcioni:2022llproc} revisited the conditions formulated by Joglekar and Lee~\cite{Joglekar:1975nu} and established a framework based on BRST symmetry, that allows inferring the gauge-variant operators and their associated renormalization factors at fixed $n$ order-by-order in the loop expansion and the strong coupling constant. They demonstrated their method with the derivation of the three-loop anomalous dimensions for $n=6$ and of the four-loop anomalous dimensions for $n$ up to 4. 
 
In this study, we propose to overcome the current lack of understanding of the all-$n$ structure of the gauge-variant operators by devising a procedure for the direct extraction of the all-$n$ Feynman rules resulting from these operators. The method and results for all-$n$ Feynman rules will be presented in section~\ref{sec:Method}. In section~\ref{sec:Application}, we apply the newly derived Feynman rules to the computation of singlet splitting functions with the massless off-shell OME method and reproduce the all-$n$ three loop results of~\cite{Vogt:2004mw}. Finally, we conclude in section~\ref{sec:conclusion}.

\section{Systematic method of deriving all-$n$ Feynman rules of gauge-variant operators}
\label{sec:Method}
We start from the generalization of Eq.~\eqref{eq:mixingOqOg} by also including the gauge-variant operators. Generally, the renormalization of the physical quark and gluon operators $O_q$ and $O_g$ can be written as the following form, 
 \begin{eqnarray}
O^{\text{R}}_q &=& Z_{qq} O^{\text{B}}_q + Z_{qg} O^{\text{B}}_g + \sum_{i=1}^\infty \sum_{j=1}^{N_i}
\left( Z_{qA_{i,j}} O^{\text{B}}_{A_{i,j}} + Z_{qB_{i,j}} O^{\text{B}}_{B_{i,j}} + Z_{qC_{i,j}} O^{\text{B}}_{C_{i,j}}\right)\,, 
\label{eq:RenorGVOq} \\
O^{\text{R}}_g &=& Z_{gq} O^{\text{B}}_q + Z_{gg} O^{\text{B}}_g + \sum_{i=1}^\infty \sum_{j=1}^{N_i}
\left(  Z_{gA_{i,j}} O^{\text{B}}_{A_{i,j}} + Z_{gB_{i,j}} O^{\text{B}}_{B_{i,j}} + Z_{gC_{i,j}} O^{\text{B}}_{C_{i,j}}\right)\,,
\label{eq:RenorGV}
\end{eqnarray}  
where the gauge-variant operators are denoted as $O_{A_{i,j}}$ for operators involving only gluon fields, $O_{B_{i,j}}$ involving two quarks plus gluons
and $O_{C_{i,j}}$ involving two ghosts plus gluons.
We stress that, as for $O_q$ and $O_g$, the gauge-variant operators are considered for generic $n$. 
The index $i$ for the operators is assigned such that the renormalization constants $Z_{qA_i}\,,Z_{qB_i}\,,Z_{qC_i}$ and $Z_{gA_i}\,,Z_{gB_i}\,,Z_{gC_i}$ are starting at ${\cal O} (\alpha_s^{i+1})$ and ${\cal O} (\alpha_s^{i})$, respectively.
The index $i$ is taken to extend to infinity since the twist-two operators are not renormalizable in the sense that they have infinite mass dimension when $n$ tends to infinity. However, only a finite number of gauge variant operators are needed at finite order of $\alpha_s$.
We note that Joglekar and Lee have shown~\cite{Joglekar:1975nu}, that the renormalization of the gauge-variant operators does not involve the physical operators.

Our key idea is to derive the (counter-term) Feynman rules due to the gauge-variant operators instead of trying to derive the operators themselves. The main ingredient for deriving the Feynman rules is to consider the one-particle-irreducible (1PI) off-shell OMEs with multiple legs for both sides of Eqs.~\eqref{eq:RenorGVOq} and \eqref{eq:RenorGV}, and impose the renormalization conditions. 
We only consider Eq.~\eqref{eq:RenorGV} in the following, since we find it to be sufficient to derive the desired Feynman rules.
Since the OMEs of the renormalized operator $O_g^{\text{R}}$ must be finite, if the divergences between the OMEs of the operators $O_q^{\text{B}}$ and $O_g^{\text{B}}$ together with corresponding renormalization factors do not cancel each other, contributions from the gauge-variant operators are needed to render the right-hand side of the equation finite.
\begin{figure}
\begin{minipage}{5cm}
\includegraphics[scale=1.0]{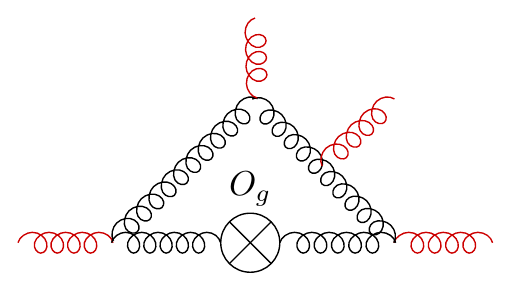} 
\end{minipage}
\begin{minipage}{5cm}
\includegraphics[scale=1.0]{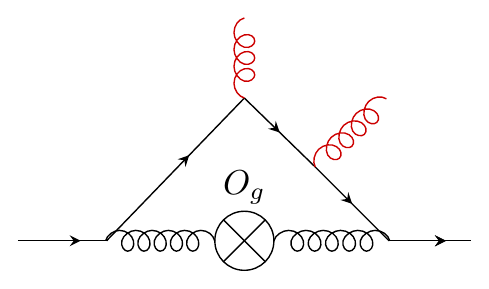} \end{minipage}
\begin{minipage}{5cm}
\includegraphics[scale=1.0]{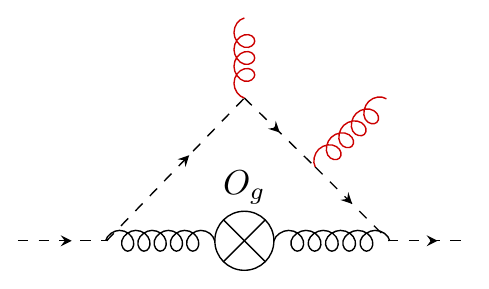} \end{minipage}
\caption{Sample Feynman diagrams for the direct extraction of Feynman rules for the 4$g$ vertex from $O_{A_1}$ (left), $q \bar{q} g g$ vertex from $O_{B_1}$ (middle) and $c \bar{c} g g$ vertex from $O_{C_1}$ (right). All diagrams involve an insertion of the physical operator $O_g$.}
\end{figure}

Let us focus on the contributions with $i=1$ in Eqs.~\eqref{eq:RenorGVOq} and \eqref{eq:RenorGV} due to the operators $A_{1,j}$, $B_{1,j}$ and $C_{1,j}$.
We will see that it is sufficient to assume that there is only one gauge-variant operator for each type of involved fields, that is, $N_1=1$.
Furthermore, we identify the renormalization constants
\begin{align}
\label{eq:identicalRC}
Z_{qA_{1,1}} &= Z_{qB_{1,1}} = Z_{qC_{1,1}} \equiv \kappa_1\,,\\
Z_{gA_{1,1}} &= Z_{gB_{1,1}} = Z_{gC_{1,1}} \equiv \eta_1\,,
\label{eq:identicalRC1}
\end{align}
which allows us to consider the combination
\begin{equation}
    O_{A_{1}}+O_{B_{1}}+O_{C_{1}}\,,
\end{equation}
where we abbreviated
$O_{A_{1}}\equiv O_{A_{1,1}}$,
$O_{B_{1}}\equiv O_{B_{1,1}}$, and
$O_{C_{1}}\equiv O_{C_{1,1}}$.
We find in our explicit calculations that Eqs.~\eqref{eq:identicalRC} and \eqref{eq:identicalRC1} are compatible with the requirement of transversity of the gluon field. With this, the renormalization of the physical operators reads
 \begin{eqnarray}
O^{\text{R}}_q &=& Z_{qq} O^{\text{B}}_q + Z_{qg} O^{\text{B}}_g + \kappa_1 \left( O_{A_{1}}+O_{B_{1}}+O_{C_{1}} \right) + \ldots,
\label{eq:renqABC1}
\\
O^{\text{R}}_g &=& Z_{gq} O^{\text{B}}_q + Z_{gg} O^{\text{B}}_g + \eta_1 \left( O_{A_{1}}+O_{B_{1}}+O_{C_{1}} \right) + \ldots,
\label{eq:rengABC1}
\end{eqnarray}  
where we omitted $i\geq 2$ type terms in this first preliminary study.

As an example, to determine the Feynman rules for the $i=1$ gauge-variant operators, we only need to consider the 1PI one-loop off-shell OMEs with $O_g$ insertion since $Z_{gA_{i,j}}\,,Z_{gB_{i,j}}\,,Z_{gC_{i,j}}$ start at $\mathcal{O}(\alpha_s^2)$ or higher for $i\geq 2$. More precisely, to extract the Feynman rules for the ghost operator $O_{C_1}$ at order $g_s^m$, we start from the following 1PI off-shell OMEs with $m$ gluons and a pair of ghost anti-ghost states, 
\begin{align}
\label{eq:FirstGhost}
&\braket{c|O_g|c+ m\,g}^{\mu_1\cdots\mu_m,\,\text{R}}_{\text{1PI}} =  Z_c (\sqrt{Z_A})^m  \big<c|Z_{gq} O_q + Z_{gg} O_g
+ \eta_1 O_{C_1} + \ldots  |c+m \,g \big> ^{\mu_1\cdots\mu_m,\,\text{B}}_{\text{1PI}} \,, 
\end{align}
where $Z_c$ and $Z_A$ are wave function renormalization constants for ghost and gluon fields, respectively.
We perform a double expansion according to the number of loops ($l$) and legs ($m+2$),
\begin{align}
 \braket{c|O_g|c+ m\,g}^{\mu_1\cdots\mu_m\text{}} = \sum_{l=0}^{\infty}
 \braket{c|O_g|c+ m\,g}^{\mu_1\cdots\mu_m, \,(l),\,(m)\text{}}_{\text{}}
 \left(\frac{\alpha_s}{4 \pi}\right)^l  g_s^m
 \end{align}
and similarly for the other operators.
To lowest order in $\alpha_s$, imposing the renormalization condition, it is easy to express the Feynman rules through the single pole part of the 1PI one-loop OMEs,  
\begin{align}
\label{eq:FirstGhostFey}
\textcolor{black}{{\eta^{(1)}_1 \braket{c|O_{C_1}|c+ m\,g}^{\mu_1\cdots\mu_m, \,{(0)},\,(m)}_{\text{1PI}}=- \left[\braket{c|O_g|c+ m\,g}^{\mu_1\cdots\mu_m, \,{(1)},\,(m),\,\text{B}}_{\text{1PI}} \right]_{1/\epsilon}  }} \,. 
\end{align}
Here, $\eta_1^{(1)}$ is defined through $\eta_1 = \alpha_s/(4 \pi) \, \eta_1^{(1)} + \mathcal{O}(\alpha_s^2)$ and can easily be determined for symbolic $n$ up to an overall constant by evaluating the right-hand-side for $m=0$ and factorizing the dependence on the kinematics. Similarly, we apply the method to the determination of Feynman rules for $O_{A_1}$ and $O_{B_1}$ and obtain
\begin{align}
\label{eq:formulaOA}
\braket{g|O_{A_1}|g+ m\,g}&^{\mu\nu\mu_1\cdots\mu_m,\,(0),\,(m)\text{}}_{\text{1PI}}  = -\frac{1}{\eta_1^{(1)}} \bigg\{  \left[ \braket{g|O_g|g+ m\,g}^{\mu\nu\mu_1\cdots\mu_m,\,(1),\,(m),\,\text{B}}_{\text{1PI}} \right]_{1/\epsilon}  \nonumber \\
& +  \left[ Z^{(1)}_{gg} - \frac{m}{2 \epsilon} \beta_0 +  \frac{m+2}{2} Z^{(1)}_A \right] \braket{g|O_g|g+ m\,g}^{\mu\nu\mu_1\cdots\mu_m,\,(0),\,(m)\text{}}_{\text{1PI}}   \bigg\}\,,  
\end{align}
\begin{align}
\label{eq:formulaOB}
\braket{q|O_{B_1}|q+ m\,g}&^{\mu_1\cdots\mu_m, \,(0),\,(m)}_{\text{1PI}} = - \frac{1}{\eta^{(1)}_1}\bigg\{ Z^{(1)}_{gq}  \left[\braket{q|O_q|q+ m\,g}^{\mu_1\cdots\mu_m, \,(0),\,(m)\text{}}_{\text{1PI}} \right] & \nonumber \\
 &+  \left[\braket{q|O_g|q+ m\,g}^{\mu_1\cdots\mu_m, \,(1),\,(m),\,\text{B}}_{\text{1PI}} \right]_{1/\epsilon} \bigg\}
  \,.
\end{align}
Equations~\eqref{eq:FirstGhostFey}, \eqref{eq:formulaOA} and \eqref{eq:formulaOB} translate the task of finding the Feynman rules for the $i=1$ type gauge-variant operators into the computations of 1PI one-loop multi-point off-shell OMEs. The details of these computations will be presented in a forthcoming paper. The extraction of contributions from $i \geq 2$ type gauge-variant operators requires us to consider the multi-point, off-shell OMEs at two loops or higher, which is in itself a complicated task and still work in progress. However, the Feynman rules up to $g_s^2$ for the $i=1$ type gauge-variant operators are proving to be sufficient to extract the three-loop, all-$n$, singlet splitting functions in the Feynman gauge, which will be presented in the next section. 

The Feynman rules to order $g_s$ for $O_{A_1}$ and $O_{C_1}$ were first listed in the reference~\cite{Hamberg:1991qt}. Upon correction of some typos there~\cite{Blumlein:2022ndg}, we find full agreement with them. The Feynman rules to order $g_s$ for $O_{B_1}$ were first given in~\cite{Matiounine:1998ky}, we also find full agreement. Our Feynman rules at order $g_s^2$ for the $i=1$ type of gauge-variant operators are new.
We present them in the following, with the convention of all momenta flowing into the vertex. The Feynman rules due to $O_{B_1}$ for quark interactions with up to 2 additional gluons are obtained as
\begin{align}
\label{eq:OB1Feynmanrules}
&\includegraphics[scale=1.0]{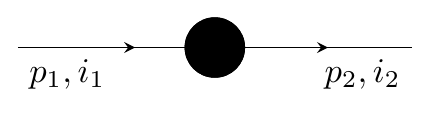}
\hspace{\textwidth}
\nonumber\\
&\quad \to 0  \,, \\[.5ex]
&\includegraphics[scale=1.0]{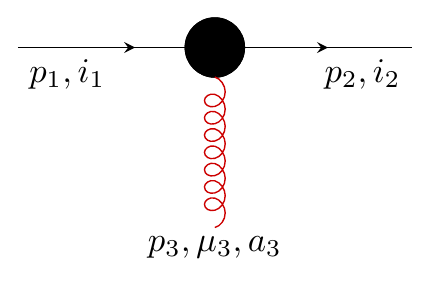}\nonumber\\
&\quad \to -\frac{1}{2} \frac{(1+(-1)^n)}{2 } \,i \,g_s  \, \Delta ^{\mu _3}  T^{a_3}_{i_2 i_1} \slashed{\Delta}_{} \big(\Delta
   \cdot \left(p_1+p_2\right)\big){}^{n-2} \,, \\[.5ex]
& \includegraphics[scale=1.0]{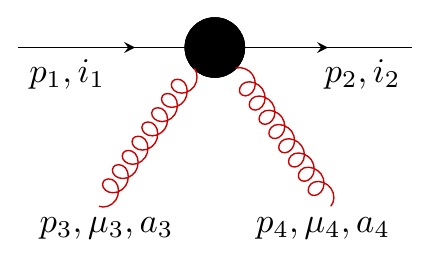}\nonumber\\
&\quad \to -\frac{i}{8} \frac{(1+(-1)^n)}{2 } g_s^2 \Delta ^{\mu _3} \Delta ^{\mu _4}
   \left( T^{a_3} T^{a_4} -   T^{a_4} T^{a_3}  \right)_{i_2 i_1} \slashed{\Delta}_{}
\nonumber\\ &\quad\quad
     \sum_{j_1=0}^{n-3}  \bigg(
     3 \left(\Delta \cdot \left(p_1+p_2\right)\right){}^{-j_1+n-3} \big[  \left(-\Delta \cdot p_3\right){}^{j_1}  - \left(-\Delta \cdot p_4\right){}^{j_1} \big] -\left(-\Delta \cdot p_4\right){}^{j_1}
   \left(\Delta \cdot p_3\right){}^{-j_1+n-3} \bigg) \,.  
\end{align}
The Feynman rules due to $O_{A_1}$ for the interaction of up to 4 gluons are obtained as
\begin{align}
&\includegraphics[scale=1]{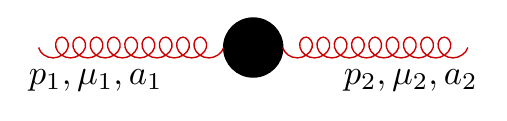} \nonumber \hspace{\textwidth}
\\
\label{eq:OA1FeynmanRules}
&\quad \to -  \delta^{a_1 a_2} \frac{(1+(-1)^n)}{2} \bigg[-\Delta \cdot p_1 \left(p_1^\mu \Delta^\nu + \Delta^\mu p_1^\nu  \right) + 2 \Delta^\mu \Delta^\nu p_1^2 \bigg] (\Delta \cdot p_1)^{n-2}  \,,
\\[0.5ex]
&\includegraphics[scale=1.0]{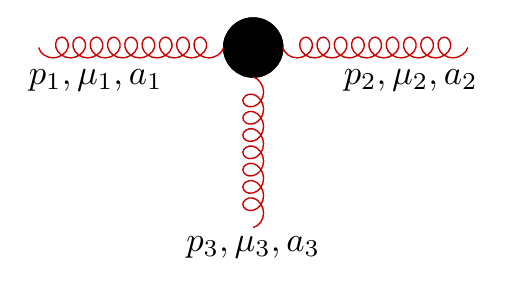} \nonumber \\
& \quad \to  \frac{1}{8} \frac{(1+(-1)^n)}{2 } g_s f^{a_1 a_2 a_3} \bigg(-4 \Delta ^{\mu _3} g^{\mu _1 \mu _2} \Delta \cdot p_1 \left(\Delta \cdot
   \left(p_1+p_2\right)\right){}^{n-2}
\nonumber\\ &\quad
   -3 \Delta ^{\mu _1} \Delta ^{\mu _3} p_2^{\mu _2} \sum_{j_1=0}^{n-2}\left(\left(-\Delta \cdot
   p_2\right){}^{j_1} \left(\Delta \cdot p_1\right){}^{-j_1+n-2}\right)+2 \Delta ^{\mu _1} \Delta ^{\mu _2} \left(4
   p_2^{\mu _3}+p_3^{\mu _3}\right) \left(\Delta \cdot p_1\right){}^{n-2}
\nonumber \\& \quad   
   -\Delta ^{\mu _1} \Delta ^{\mu _2} \Delta ^{\mu _3} \left(p_1\cdot p_1-p_1\cdot
   p_2+p_2\cdot p_2\right) \sum_{j_1=0}^{n-3}\left(\left(-\Delta \cdot p_2\right){}^{j_1} \left(\Delta \cdot
   p_1\right){}^{-j_1+n-3}\right) \bigg) + \text{{\it{permutations}}} \,,
\\[0.5ex]
&\includegraphics[scale=1.0]{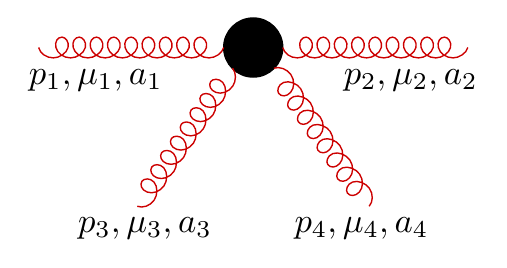}
\nonumber \\
& \to \frac{1}{96}\frac{(1+(-1)^n)}{2 }  i g_s^2 \bigg\{3 \Delta ^{\mu _1} \Delta ^{\mu _2} g^{\mu _3 \mu _4} \bigg(8 \left(2 f^{a a_1 a_3}
   f^{a a_2 a_4} -f^{a a_1 a_2} f^{a a_3 a_4} \right) \left(\Delta \cdot
   p_1\right){}^{n-2}  \nonumber 
   \\
   &\quad -f^{a a_1 a_2} f^{a a_3 a_4} \left(\Delta \cdot p_1+\Delta \cdot p_2+2 \Delta
   \cdot p_3\right)  \nonumber 
   \\
   &\quad  \times \sum_{j_1=0}^{n-3} \bigg[ \big[6  \left(\Delta \cdot \left(-p_1-p_2\right)\right){}^{j_1}+\left(\Delta \cdot p_2\right){}^{j_1} \big] \left(-\Delta \cdot
   p_1\right){}^{-j_1+n-3}  \bigg] \bigg)\nonumber 
   \\
   &\quad 
   +2 \Delta ^{\mu _1} \Delta ^{\mu _2} \Delta ^{\mu _3} \Delta ^{\mu _4}
   \bigg(\left(p_1\cdot p_1+p_1\cdot p_2+p_1\cdot p_3+p_2\cdot p_2+p_2\cdot p_3+p_3\cdot p_3\right)
   f^{a a_1 a_3} f^{a a_2 a_4} \nonumber 
   \\
   &\quad+\left(4 p_1\cdot p_1-5 p_1\cdot p_2-5 p_1\cdot p_3-5 p_2\cdot p_2-5
   p_2\cdot p_3+4 p_3\cdot p_3\right) f^{a a_1 a_2} f^{a a_3 a_4} \bigg)
   \nonumber 
   \\
   &\quad \times \sum_{j_1=0}^{n-4} \sum_{j_2=0}^{j_1} \left(\left(-\Delta
   \cdot p_3\right){}^{j_2} \left(\Delta \cdot \left(p_1+p_2\right)\right){}^{j_1-j_2} \left(\Delta \cdot p_1\right){}^{-j_1+n-4}\right) \nonumber 
   \\
   &\quad
   -\Delta ^{\mu _2} \Delta ^{\mu _4} \left( f^{a a_1 a_3}
   f^{a a_2 a_4}+13 f^{a a_1 a_2} f^{a a_3 a_4} \right) \left(\Delta ^{\mu _1} p_3^{\mu
   _3}-\Delta ^{\mu _3} p_1^{\mu _1}\right)\nonumber 
   \\
   &\quad\times \sum_{j_1=0}^{n-3} \sum_{j_2=0}^{j_1}  \left(\left(\Delta \cdot \left(-p_1-p_2\right)\right){}^{j_1-j_2}
   \left(\Delta \cdot p_3\right){}^{j_2} \left(-\Delta \cdot p_1\right){}^{-j_1+n-3}\right)
   +3 \Delta ^{\mu _1} \Delta^{\mu _2} \Delta ^{\mu _4}\nonumber 
   \\
   &\quad \times \left(4 p_1^{\mu _3}+p_3^{\mu _3}\right) f^{a a_1 a_3} f^{a a_2 a_4}
   \sum_{j_1=0}^{n-3} \bigg[ \big[  4 \left(\Delta \cdot \left(p_1+p_3\right)\right){}^{j_1} + \left(\Delta \cdot p_4\right){}^{j_1} \big]  \left(-\Delta \cdot
   p_2\right){}^{-j_1+n-3} \nonumber 
   \\
   &\quad   +2  \left(\Delta \cdot p_4\right){}^{j_1} \left(\Delta \cdot
   \left(-p_1-p_3\right)\right){}^{-j_1+n-3} \bigg] + d_r^{a_1 a_2 a_3 a_4} (\cdots)\bigg\} + \text{{\it{permutations}}} \,,
\end{align}
where plus {\it{permutations}} means that we add the contributions from permutations of all involved gluons.
Further, $d_r^{a_1 a_2 a_3 a_4}$ is the totally symmetric color structure and contributes to the splitting functions only at the four-loop level. The Feynman rules for $O_{C_1}$ and the color structure $d_r^{a_1 a_2 a_3 a_4}$ in Eq.~\eqref{eq:OA1FeynmanRules} will be presented in a future publication. Our results in Eq.~\eqref{eq:OA1FeynmanRules}, Eq.~\eqref{eq:OB1Feynmanrules} and also the Feynman rules for $O_{B_1}$ are in closed forms with generic $n$ dependence and with the renormalization constants being factored out. Therefore, it shows that only three operators, i.e.\ $O_{A_1}\,, O_{B_1}$ and $O_{C_1}$, are needed for $i=1$ and up to 4 legs, thus providing strong evidence for our statement right above Eq.~\eqref{eq:identicalRC}. It also explicitly verifies the correctness of Eqs.~\eqref{eq:identicalRC} and~\eqref{eq:identicalRC1} at the one-loop level.
As we discuss in the following, our setup allows us to successfully reproduce the literature result for the three-loop singlet splitting functions.

\section{Three-loop singlet splitting functions from off-shell OMEs}
\label{sec:Application}
With the newly derived Feynman rules for the $i=1$ type gauge-variant operators in section~\ref{sec:Method}, we are ready to apply the off-shell OME method to compute the three-loop singlet splitting functions. Unlike the derivation of Feynman rules in the last section, at this point, we only need to compute two-point multi-loop off-shell OMEs. However, one can not naively apply integration-by-part (IBP) reductions~\cite{Chetyrkin:1981qh} to the OMEs since non-standard terms involving the symbol $n$ appear in the Feynman rules of the twist-two operators, see for example Eq.~\eqref{eq:OA1FeynmanRules}. For example, the terms like $(\Delta \cdot p)^{n-2}$ with scalar products raised to arbitrary power need to be dealt with properly. We adopt the method first proposed in~\cite{Ablinger:2012qm,Ablinger:2014yaa} to sum the non-standard terms into a linear propagator with the help of a tracing parameter $x$, 
\begin{align}
\label{eq:ToLinearP}
(\Delta \cdot p)^{n-2} \to \sum_{n=2}^\infty x^n (\Delta \cdot p)^{n-2} = \frac{x^2}{ {1-x \Delta \cdot p} } \,. 
\end{align}
With the help of Eq.~\eqref{eq:ToLinearP} and its generalization to multiple sums, we are able to map all Feynman rules to quantities in resummed-$x$ space involving linear propagators.
We work in resummed-$x$ space throughout and revert back to $n$ space at the end of the calculation by expanding in $x$ and extracting the coefficient of $x^n$. 

In this way, the computation follows a standard computational framework. We use \texttt{QGRAF}~\cite{Nogueira:1991ex} to generate all relevant Feynman diagrams, and substitute the resummed-$x$ space Feynman rules in \texttt{Mathematica}. Subsequently, \texttt{FORM}~\cite{Vermaseren:2000nd} is used to to evaluate the Dirac and color algebra. Then, we use \texttt{Reduze 2}~\cite{vonManteuffel:2012np} as well as \texttt{FeynCalc}~\cite{Shtabovenko:2016sxi} to classify the list of integrals into a small set of integral families. Several public packages implementing Laporta's algorithm~\cite{Laporta:2000dsw} were used extensively to perform IBP reductions: \texttt{FIRE6}~\cite{Smirnov:2019qkx} in combination with \texttt{LiteRed}~\cite{Lee:2012cn}, \texttt{Reduze 2}~\cite{vonManteuffel:2012np} as well as \texttt{Kira}~\cite{Klappert:2020nbg}.
\begin{figure}
\begin{minipage}{5cm}
\includegraphics[scale=1.0]{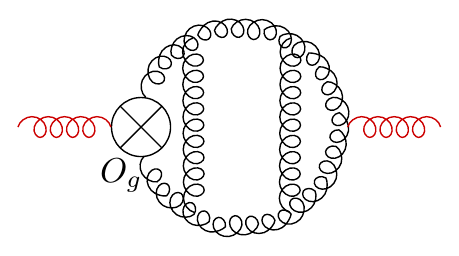} 
\end{minipage}
\begin{minipage}{5cm}
\includegraphics[scale=1.0]{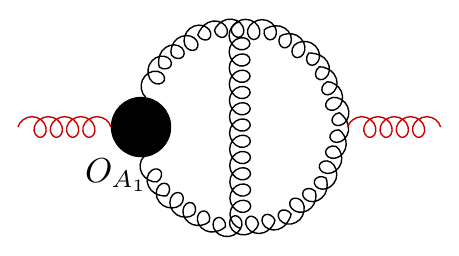} 
\end{minipage}
\begin{minipage}{5cm}
\includegraphics[scale=1.0]{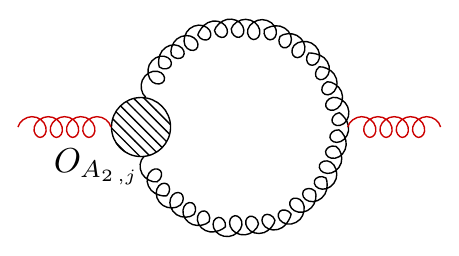} 
\end{minipage}
\caption{Sample Feynman diagrams for three-loop singlet splitting functions, with contributions from the physical operators to three loops (left), the $i=1$ type gauge-variant operators to two loops (middle) and the $i=2$ type gauge-variant operators to one loop (right).}
\end{figure}

To solve the master integrals, we choose to derive differential equations~\cite{Kotikov:1990kg,Bern:1993kr,Gehrmann:1999as} with respect to the parameter $x$~\cite{Ablinger:2015tua}.
We manage to turn them into canonical form~\cite{Henn:2013pwa} using the public packages \texttt{CANONICA}~\cite{Meyer:2017joq, Meyer:2016slj} and \texttt{Libra}~\cite{Lee:2014ioa,Lee:2020zfb}.
The integration involves only the forms $\mathrm{d}\ln(x)$, $\mathrm{d}\ln(1-x)$ and $\mathrm{d}\ln(1+x)$, and the boundary constants are conveniently fixed at $x=0$, where the integrals are just standard massless propagator integrals~\cite{Chetyrkin:1980pr,Tkachov:1981wb,Chetyrkin:1981qh,Gorishnii:1989gt,Ruijl:2017cxj}.
In this way, the off-shell OMEs are written in terms of harmonic polylogarithms~\cite{Remiddi:1999ew} with argument $x$. The OMEs in $x$ space are turned into $n$ space expressions in terms of harmonic sums~\cite{Vermaseren:1998uu,Blumlein:1998if} with the help of the Mathematica package \texttt{HarmonicSums}~\cite{Ablinger:2009ovq,Ablinger:2014rba}. In the end,  we obtain $\braket{j(p)|O_{q/g}|j(p)}$ to three loops and $\braket{g(p)|O_{A_1}|g(p)}$, $\braket{q(p)|O_{B_1}|q(p)}$, $\braket{c(p)|O_{C_1}|c(p)}$ to two loops with general $\xi$ dependence, where $\xi$ is the gauge parameter and $\xi=1$ in Feynman gauge.

From Eq.~\eqref{eq:RenorGV} it is easy to tell that the one-loop OMEs of $i=2$ type operators also contribute to the three-loop singlet splitting functions for general $\xi\neq 1$. Their associated Feynman rules are still work in progress. As a first step, we just omit the $i=2$ contributions (dots) in
Eqs.~\eqref{eq:renqABC1}-\eqref{eq:rengABC1}.
Interestingly, we still reproduce the three-loop all-$n$ singlet splitting functions
$\gamma^{(2)}_{kl}\big\vert_{\text{VMV}}$
presented in Ref.~\cite{Vogt:2004mw} in Feynman gauge with $\xi=1$, 
\begin{alignat}{2}
\label{eq:resultsWithoutO2}
\left.\gamma^{(2)}_{qq}\right\vert_{\text{this study}} - \left.\gamma^{(2)}_{qq}\right\vert_{\text{VMV}} &= 0\,,&\qquad
\left.\gamma^{(2)}_{qg}\right\vert_{\text{this study}} - \left.\gamma^{(2)}_{qg}\right\vert_{\text{VMV}}
&= 0 \,, \nonumber \\
\left.\gamma^{(2)}_{gq}\right\vert_{\text{this study}} - \left.\gamma^{(2)}_{gq}\right\vert_{\text{VMV}} &= 0\,,&
\left.\gamma^{(2)}_{gg}\right\vert_{\text{this study}} - \left.\gamma^{(2)}_{gg}\right\vert_{\text{VMV}} &= \left(1-\xi\right)\left[ \cdots \right] \,.  
\end{alignat}
The correctness of $q$ to $g$ splitting in covariant gauge with general $\xi$ dependence implies that the Feynman rule of the $q\bar{q}g$ vertex from $i=2$ type contributions should be zero. 

\section{Conclusion}
\label{sec:conclusion}
Massless off-shell operator matrix elements allow for a very efficient computation of higher-order splitting functions.
In the singlet case, however, there is a longstanding problem: the physical quark and gluon operators mix with some unknown gauge-variant operators under renormalization.
We developed a novel method to directly extract the all-$n$ Feynman rules resulting from those unknown operators. The method translates the determination of Feynman rules into the computations of multi-loop multi-point OMEs.
As a first study, we obtained the all-$n$ Feynman rules for the leading contributions of the gauge-variant operators and applied them to the computation of the three-loop singlet splitting functions. Interestingly, we found that we reproduced the three-loop all-$n$ singlet functions in Ref.~\cite{Vogt:2004mw} in Feynman gauge even though the next-to-leading contributions of the gauge-variant operators were not included.
Work on the Feynman rules associated to these contributions is still in progress. 

\acknowledgments
This work was supported in part by the Swiss National Science Foundation (SNF) under contract 200020-204200, by the European Research Council (ERC) under the European Union's Horizon 2020 research and innovation programme grant agreement 101019620 (ERC Advanced Grant TOPUP), and by the National Science Foundation (NSF) under grant 2013859.

\bibliographystyle{JHEP}
\bibliography{biblio}

\end{document}